# Transmitting Video-on-Demand Effectively

Rachit Mohan Garg
CSE Deptt, JUIT, INDIA
rachit.mohan.garg@gmail.com

Shipra Kapoor
ECE Deptt, JUIT, INDIA
ece.shiprakapoor.dit@gmail.com

Kapil Kumar
CSE Deptt, JUIT, INDIA
kapil.cs89@gmail.com

Mohd. Dilshad Ansari
CSE Deptt, JUIT, INDIA
m.dilshadcse@gmail.com

*Abstract*—Now-a-days internet has become a vast source of entertainment & new services are available in quick succession which provides entertainment to the users. One of this service i.e. Video-on-Demand is most hyped service in this context. Transferring the video over the network with less error is the main objective of the service providers. In this paper we present an algorithm for routing the video to the user in an effective manner along with a method that ensures less error rate than others.

## I. INTRODUCTION

Now-a-days internet has become a vast source of entertainment & new services are available in quick succession which provides entertainment to the users. One of this service i.e. Video-on-Demand is most hyped service in this context. Transferring the video over the network with less error is the main objective of the service providers.

In order to reduce the redundancies during the VoD process and communication we put forward a radius restrained distributed breadth first search flooding algorithm (RRDBFSF) [8]. By knowing the information of neighbor node in finite scope, this algorithm does breadth first search and selects the least number of forwarding neighbor nodes to reduce redundant information in broadcasting routing information. In scenario in RRDBFSF where each node receives one of many available video streams, we propose a network coding based cooperative repair framework to improve broadcast video quality during channel losses.

The proposed methodology is as follows:

1. Generating a MODA Framework for VoD Service
2. Clustering of the nodes in the network.
3. Estimation of the available bandwidth using packet probing.
4. Transmission using RRDBFS algorithm.
5. Unstructured Network Cooperative for repairing of data packet if in case error occurs.

## II. MODA FRAMEWORK

The objective of a MODA [7] is to shift the complexity from the implementation of an application to its specification. It specifies three levels of architecture:

- *Computation Independent Model* (CIM): CIM describes the context in which the system will be used.
- *Platform Independent Model* (PIM): It describes the system itself without any details of its use or its platform.
- *Platform Specific Model* (PSM): At this level environment of implementation platforms or languages is known.

### A. MODA Process

Figure 2 examines many of the ways translation could be used for Ontology Driven Architecture. The repeatable process of this design is that the 'System Translator Program' created in Step 1 creates a new 'System Translator Program' in Step 2 and this creates Visualization. The program is translated to a style diagram using a second stage of translation. The second 'System Translator Program' could also create a 'Model/Program'. 'Meta Program' or translate to an 'External Application'.

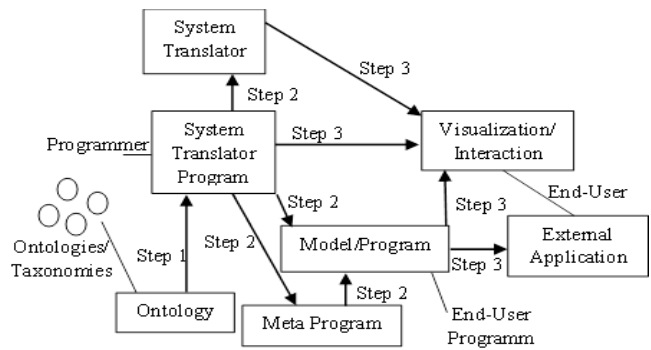

Fig. 2 MODA Process

### B. MODA Framework for Video-on-Demand Service

The MODA process (CIM to PIM) allows generating from the sending communication task a SCA domain including both server and client composites. For each composite the SessionController and MediaController components are inferred. Following the general MODA process the adequate session controller as well as the media controller implementations has been selected. XSL templates used to implement the mapping rules the MODA engines. Figure 3 illustrates this.

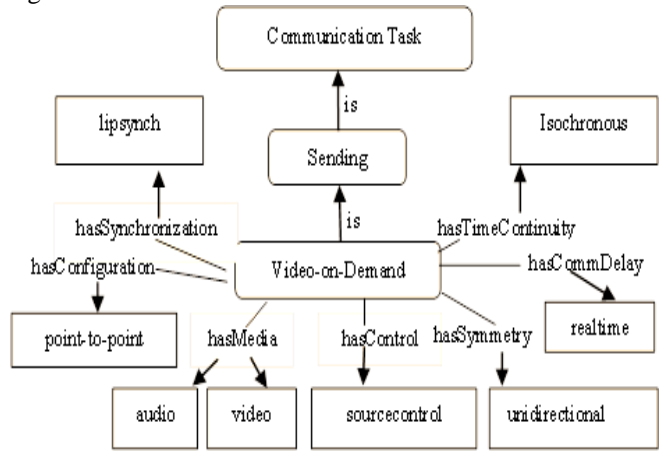

Fig. 3 VoD Process with MODA Framework

Following the MODA process (PIM to PSM), the adequate session controller (i.e. RTSP server and client) as well as the media controller implementations (i.e. JMF media controllers) have been selected. Figure 4 depicts the PSM SCA individual for the VoD system.

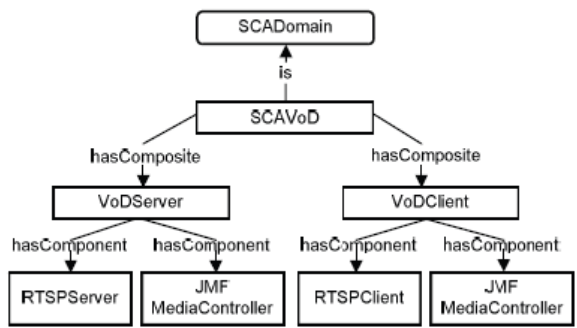

Fig. 4 SCA Description of VoD System

### III. CLUSTERING OF NODES IN THE NETWORK

When a node joins network, it sets the cluster state to INITIAL. Moreover, the state of a floating node (a node does not belong to a cluster yet) also sets to INITIAL. Because passive clustering exploits on-going packets, the implementation of passive clustering resides between layer 3 and 4. The IP option field for cluster information is as follows:

Node ID: The IP address of the sender node. This is different to the source address of the IP packet.

State of cluster: The cluster state of the sender node.

If a sender node is a gateway, then it tags two IP addresses of cluster heads (CHs) which are reachable from the gateway. We summarize the passive clustering algorithm as follows:

*Cluster states*: There are 6 possible states; INITIAL, CLUSTER HEAD, ORDINARY NODE, GATEWAY, CH READY, GW READY and DIST GW.

The packet handling upon sending a packet, each node piggybacks cluster-related information. Upon a promiscuous packet reception, each node extracts cluster-related information of neighbors and updates neighbor information table.

*a) cluster head (CH) declaration*

A node in INITIAL state changes its state to CH READY (a candidate cluster head) when a packet arrives from another node that is not a cluster head. With outgoing packet, a CH READY node can declare as a cluster head (CH). This helps the connectivity because this reduces isolated clusters.

*b) Becoming a member*

A node becomes a member of a cluster once it has heard or overheard a message from any cluster head. A member node can serve as a gateway or an ordinary node depending on the collected neighbor information. A member node can settle as an ordinary node only after it has learned enough neighbor gateways. In passive clustering, however, the existence of a gateway can be found only through overhearing a packet from that gateway. Thus, we define another internal state, GW READY, for a candidate gateway node that has not yet discovered enough neighbor gateways. Recall that we develop a gateway selection mechanism to reduce the total gateways in the network. The detailed mechanism will be shown in the next Section. A candidate gateway finalizes its role as a gateway upon sending a packet (announcing the gateway's role). Note that a candidate gateway node can become an ordinary node any time with the detection of enough gateways.

### IV. BANDWIDTH ESTIMATION BEFORE TRANSMISSION

When the service providers receive a request for a video from a user than a dummy packet is sent along the paths that lead to the requester. On receiving the destination packets reverts it back to the source i.e. the service provider. This dummy packet is responsible for the estimation of the characteristics of the link. At the service providers side all the dummy packets are analyzed and the effective bandwidth of the user is estimated. According to the effective bandwidth of the channel the original video is than compressed according to the channel's bandwidth for the ease of transmission. This method is known as *Packet Probing* [3].

After the estimation & compression the video is transmitted according to RRDBFSF algorithm which is described in the next section.

*Corresponding Author: Rachit Mohan Garg, CSE Deptt, JUIT, INDIA*

## V. Radius Restrained Distributed Breadth First Search Flooding Algorithm (RRDBFSF) For Video On Demand (VoD) Service

Our algorithm is flooding in a small scope, namely, radius restrained flooding algorithm, and can reduce redundancies within a certain scope. Based on the rule that lessen the cost and time of forwarding message between nodes to the best, we choose the scope within a radius of three to flood message. So, every node need know its neighbor nodes which are connected directly with it, and need realize some information about their neighbor nodes. We call these information are nodes information within a radius of three. Thus, we suppose that $v$ is random node in the networks.

### A. Description of RRDBFSF Algorithm

*1) The Rules followed in the Algorithm:*

- The least cost of forwarding message time.
- Node is only concerned about the nodes flooding within a radius of three.
- When there are more than one path that message can get to the destination, we choose the shortest path.

*2) Given Conditions of RRDBFSF Algorithm:*

- The network is connected entirely ensuring that there is a reachable path between any two nodes.
- Any connection between nodes is bidirectional.
- Before running the flooding algorithm, the node has received its neighbor nodes information and built the neighbor nodes table NT (v).

*3) Some Common Terms:*

- Node Set

TABLE I.  Node Set which defines

| Notation | Description |
|---|---|
| N(x) | The neighbor nodes set of node x. |
| RN(x) | The relative neighbor nodes set of node x, viz. RN (x). |
| N (x) | The set is calculated by the RRDBFSF algorithm. |
| TLen(x) | The neighbor nodes set of node x within a radius of three. |
| RTLen(x) | The relative neighbor nodes set of node x within a radius of three, viz. RTLen (x). |
| TLen (x) | The set is calculated by the RRDBFSF algorithm and message from node is forwarded for the second time to the nodes in the set |
| T(x) | The sum set of node x neighbor nodes, viz. the sum of neighbor nodes and the nodes within a radius of three |
| R(x) | The set of node x next forwarding nodes |

- Neighbor Node table

*Corresponding Author: Rachit Mohan Garg, CSE Deptt, JUIT, INDIA*

NT (v)- Founded on the messages from neighbor nodes and its content is flashed real time.

RT (v)- Includes the node $v$ next forwarding neighbor nodes and their respective next forwarding neighbor nodes.

TABLE II.  Neighbor Node Set Nt(v) which defines

| Neighbor node ID | Its neighbor nodes within a radius of three |
|---|---|

TABLE III.  Neighbor Node Set Nt(v) which defines

| Next forwarding node ID | Next forwarding nodes |
|---|---|

The RRDBFSF algorithm calculates, gets RT(v), and transmits it to next forwarding neighbor nodes. So, if a forwarding node ID is $i$ in RT(v), its next forwarding nodes set concerned with node $v$ is RT (v, i).

Consider a sample network in figure 5. All the values are calculated considering $v$ as reference node.

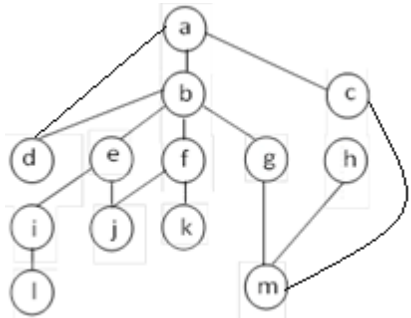

Fig. 5 A Sample Network

N (f) = {b, k, j}

TLen (f) = {c, i, m, j, b, d, a}

T (e) = {a, b, c, d, i, j, k, m}

TABLE IV.  Sample Of Neighbor Nodes Table

| Neighbor Node ID | Neighbor Node in radius of 3 |
|---|---|
| b | f, g, h, l, c |
| k | e, g, a, d |
| j | a, g, d, f, e, l |

## VI. Network Based Coding

### A. System Architecture For Video Repair System in VoD

*1) CPR System Architecture:* We consider the scenario where N peers are watching broadcasting video streams through their wireless mobile devices and mobile devices are equipped with wireless local area network (WLAN) interfaces. We first assume that the media source provides a total of Sall video streams. Sall varies due to different technologies, broadcast technologies, broadcast bandwidths and operational constraints of the mobile video providers. Although Sall stream are available, not all streams will have audiences in a given ad-hoc network at a given time. Without loss of generality, we denote $S^* = \{s1, s2, \ldots, ss\}$ as the subset of Sall streams that have audience and $S=|S^*|$. Each peer n in the network watches one stream $S(n)$ € $S^*$ from the media source and conversely each stream s € $S^*$ has a group of receiving peers $\mu_s$. Peers in $\mu_s$, each receiving a different subset of packets of stream s, can relay packets to others WLAN interfaces to repair lost packets. This repair process is called CPR i.e. Cooperative Peer to Peer Repair. We denote as the set of streams of which n peer has received packets: either original video packets from the media source or CPR packets from peers, i.e., streams that peer n can repair via CPR. We use flags in CPR packet header to identify the stream a packet repairs. Whenever n peer has a transmission opportunity - a moment in time when peer n is permitted by a scheduling protocol to locally broadcast a packet via WLAN, peer n selects one stream from An to construct and transmit a CPR packet.

*2) Source Model:* We use H.264 codec for video source encoding because of its excellent rate-distortion performance. For improved error resilience, we assume the media source first performs reference frame selection for each group of picture (GOP) in each stream separately during H.264 encoding. In brief assumes each GOP is composed of a starting I-frame followed by P-frames. Each P-frame can choose among a set of previous frames for motion compensation, where each choice results in a different encoding rate and different dependency structure. If we then assume that a frame is correctly decoded only if it is correctly received and the frame it referenced is correctly decoded probability. Figure 6 illustrates the use of DAG Model for refernce frame selection.

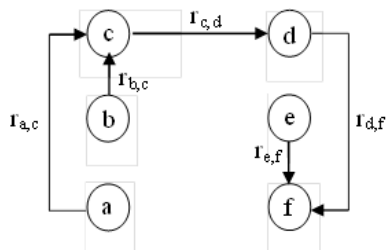

Fig. 6 Example of DAG source model for H.264/AVC video with reference frame selection

After the media source performs reference frame selection for each HOP of each stream, we can model $M^k$ frames in GOP of a stream $s^k$, $F^k = \{F_1^k \ldots F_M^k\}$, as nodes in a directed acyclic graph (DAG) in fig 4. Each frame $F_i^k$ has an associated $d_i^k$, the resulting distortion reduction if $F_i^k$ is correctly decoded. Each frame $F_i^k$ points to the frame in the same GOP that it uses for motion compensation. Frame $F_i^k$ referencing frame $F_j^k$ is packetized into real-time transport protocol (RTP) packets according to the frame size and maximum transport unit (MTU) of the delivery network. A frame $F_j^k$ is correctly received only if all packets within $F_j^k$ are correctly received. We assume that the media source delivers each GOP of $M^k$ frames of stream $s^k$ in time duration $Y^k$. $Y^k$ is also the repair epoch for $s_k$, which is the duration in which CPR completes its repair on the previous GOP, i.e peers exchange CPR packets for previous GOP of stream $s_k$ during the current epoch.

*3) CPR-Unstructured Network Coding (UNC):* We denote the traditional random NC scheme as UNC. First, suppose n peer has a transmission opportunity and n selects stream s from $A_n$ for transmission. Suppose there are M original (native) frames $F = \{F_1, \ldots F_M\}$ in a GOP of stream s to be repaired among peers in $\mu_s$. Each frame $F_i$ is divided into multiple packets $P_i = \{p_{i,1}, p_{i,2}, \ldots p_{i,Bi}\}$ of size W bits each. Here $B_i$ is the number of packets frame $F_i$ is divided into. A peer adds padding bits to each packet so that each has constant size W bits; this is performed for NC purposes. We denote $P^*$ as the set of all packets in a GOP, i.e $P^* = \{P_1, \ldots, P_M\}$. There are a total of $P=|P^*|=£_{i=1}^M B_i$ packets to be disseminated among peers in $\mu_s$.

We denote $G_n$ as the set of *native packets* of stream $S(n)$ peer 'n' received from media source. Denote $Q_n$ as the set of *NC packets* of stream 's' peer 'n' received from other peers through CPR. If the stream selected for transmission is the same as the stream peer 'n' currently watches, i.e. $s=S(n)$ then the NC packet $q_n$ generated by peer n is represented as:

$$q = \sum_{p_{i,j} \in G_n} a_{i,j} p_{i,j} + \sum_{q_m \in Q_n} b_m q_m = \sum_{p_{i,j} \in P^*} c_{i,j} p_{i,j} \quad (1)$$

where $a_{i,j}$'s and $b_m$'s, random numbers in , are coefficients for the original packets and the received encoded NC packets, respectively. Because each received NC packet $q_m$ is itself a linear combination of native and NC packets, we can rewrite $q_n$ as a linear combination of native packets with *native coefficients* $c_{i,j}$ 's.

If the stream selected for transmission $s \neq S(n)$, then the NC packet is simply a linear combination of all NC packets of stream *s* received through CPR from other peers so far as follows:

$$q_n = \sum_{q_m \in Q_n} b_m q_m = \sum_{p_{i,j} \in P^*} c_{i,j} p_{i,j} \quad (2)$$

For UNC, *all* packets of stream 's', both native packets (if any) and received NC packets, are used for NC encoding, and a peer in $\mu_s$ can reconstruct all native packets of stream when innovative native or NC packets of stream 's' are received, and hence all frames can be recovered.

## VII. RESULTS

Methodology has been simulated on a network with 10 soft switch nodes, 50 soft switch nodes, or 100 soft switch nodes.

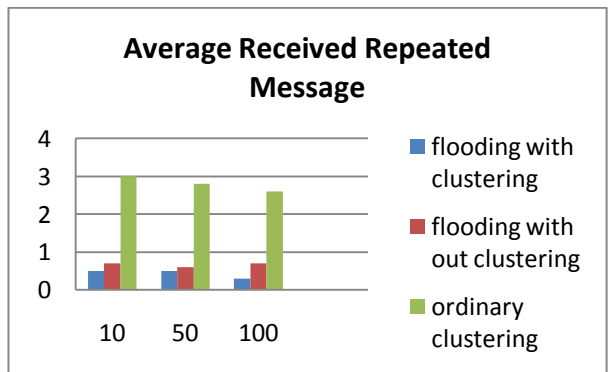

Fig. 7 Average Receive Repeated Message

**Network Scale**

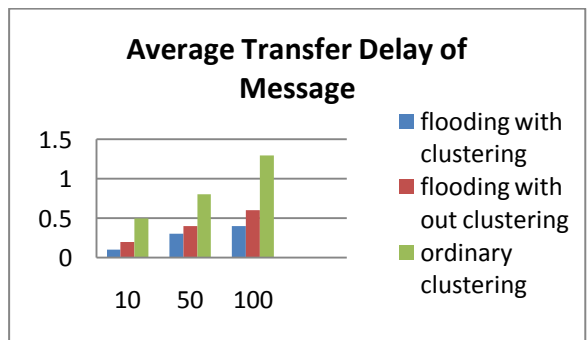

Fig. 8 Average Received repeated message in different network scale

## VIII. CONCLUSION AND FUTURE WORK

This paper presents a framework for the VoD. A cluster of the network is created & based on the estimation by *packet probing* method the data i.e. the video is sent to the destination using Radius Restrained Breadth First Search which decreases the redundant information in the network & stream the video data effectively to the destination node. To avoid any error or to ensure reliable data delivery a repair mechanism for the VoD service has been given.

The future work includes designing a more efficient algorithm which will route the video data efficiently & securely over the network. The other area of research is designing a framework for packet repairing which incurs less overhead on the basis of time & cost. Next steps in MODA framework aim at generating dynamic user interfaces for final users (to drive the deployment process) and enhancing deployment specifications produced by MODA in order to be used for self-configuring the communication system.

REFERENCES

[1] X. Liu, G. Cheung, and C.-N. Chuah, "Rate-distortion optimized network coding for cooperative video stream repair in wireless peer-to-peer networks," 1st IEEE Workshop on Mobile Video Delivery, Newport Beach, CA, June 2008.

[2] X. Liu, S. Raza, C.-N. Chuah, and G. Cheung, "Network coding based cooperative peer-to-peer repair in wireless ad-hoc networks," IEEE International Conference on Communications, Beijing, China, May 2008.

[3] Mesut Ali Ergin, Marco Gruteser, "Using Packet Probes for Available Bandwidth Estimation: A Wireless Testbed Experience," ACM, WiNTECH'06, September 29, 2006, Los Angeles, California, USA.

[4] H. Seferoglu and A. Markopoulou, "Opportunistic network coding for video streaming over wireless," 16th IEEE International Packet Video Workshop, Lausanne, Switzerland, November 2007.

[5] D. Nguyen, T. Nguyen, and X. Yang, "Multimedia wireless transmission with network coding," 16th IEEE International Packet Video Workshop, pp. 326--335, Lausanne, Switzerland, November 2007.

[6] Jorge Gómez-Montalvo, Myriam Lamolle, Ernesto Exposito, "A Multimedia Ontology Driven Architecture framework (MODA) for Networked Multimedia Systems," 1st IEEE International Conference on Networked Digital Technologies, pp. 411—416, Czech Republic, 2009.

[7] Multimedia Ontology Driven Architecture framework, http://www.laas.fr/~eexposit/pmwiki/pmwiki.php/MODA

[8] Rui Li, Shaoming Pan, Hao Wang, "Radius Restrained Distributed Breadth First Search Flooding in Large-scale Multimedia Communication," IEEE Asia-Pacific Conference on Information Processing, 2009.

[9] Andy Yoo, Edmond Chow, Keith Henderson, William McLendon., "A Scalable Distributed Parallel Breadth-First Search Algorithm on BlueGene/L," Proc of 2005 ACM/IEEE SC05 Conference, 2005.

[10] K. Nguyen, T. Nguyen, and S.-C. Cheung, "Video streaming with network coding," Springer J. Signal Process. Syst Special Issue: ICME07, February 2008.

[11] J. Crowcroft and K. Paliwoda, "A multicast transport protocol," Proc. ACM SIGCOMM, New York, August 1988.

[12] S. Raza, D. Li, C.-N. Chuah, and G. Cheung, "Cooperative peer-to-peer repair for wireless multimedia broadcast," Proc. IEEE Int. Conf. Multimedia and Expo (ICME), pp. 1075—1078, Beijing, China. July 2007.

[13] T. Wiegand, G. Sullivan, G. Bjontegaard, and A. Luthra, "Overview of the H.264/AVC video coding standard," IEEE Trans. Circuits Syst. Video Technology, vol. 13, pp. 560—576, July 2003.

*Corresponding Author: Rachit Mohan Garg, CSE Deptt, JUIT, INDIA*